\documentclass[10pt,twoside]{Lowx2011}
\usepackage{epsf,amsmath}
\usepackage{graphicx}

\begin{document}

\title{NEW LOOK AT GEOMETRICAL SCALING}

\author{\underline{MICHAL\ PRASZALOWICZ}\\ \\
M. Smoluchowski Institute of Physics\\
Jagellonian University\\
Reymonta 4, 30-059 Krakow,
Poland\\
E-mail: michalif.uj.edu.pl}

\maketitle

\begin{abstract}
\noindent In this note we discuss emergence of geometrical scaling
(firstly proposed for deep inelastic collisions) in pp scattering
at the LHC and in heavy ion collisions at RHIC. After discussing
general properties of geometrical scaling (GS) we focus on simple
signatures of GS, namely on ratios of the $p_{\rm T}$ values
at which multiplicity spectra for different energies are
equal.
\end{abstract}

\markboth{\large \sl \hspace*{0.25cm}\underline{MICHAL PRASZALOWICZ}
\hspace*{0.25cm} Low-$x$ Meeting 2011}
{\large \sl \hspace*{0.25cm} NEW LOOK AT GEOMETRICAL SCALING}

\section{Geometrical Scaling in Deep Inelastic Scattering}

One of the outstanding problems of the evolution of parton densities is the rapid increase of the number of gluons at low Bjorken $x$ \cite{HERAcombined}. Such growth has to
be tamed at some point.  The scale at which this happens is called saturation scale
$Q_{\text{s}}$ and it depends on the Bjorken $x$. The explicit form of the saturation
scale follows from the fact that
$Q_{\text{s}}^{2}(x)$ is related to the gluon distribution in the proton at low $x$
\cite{GolecBiernat:1998js}:%
\begin{equation}
Q_{\text{s}}^{2}(x)=Q_{0}^{2}\left( {x}/{x_0}\right)  ^{-\lambda}%
\label{defQsat}
\end{equation}
where $Q_{0}\sim1$ GeV and $x_{0}\sim10^{-3}$ are free parameters whose
precise values can be extracted from fits to the HERA data.
Power $\lambda$ is known to be of the order $\lambda\sim0.2\div0.3$.

Based on the saturation idea
Golec-Biernat and W\"{u}sthoff proposed a model  \cite{GolecBiernat:1998js}
where the reduced
cross-section $\sigma_{\gamma^{\ast}p} \sim F_2/Q^2$ is written as:
\begin{equation}
\sigma_{\gamma^{\ast}p}=\int dr^{2}\left\vert \psi(r,Q^{2})\right\vert
^{2}\sigma_{dP}(r^{2},x) \label{redsig}%
\end{equation}
where $\psi$ is an amplitude for photon dissociation into a $\bar{q}q$ pair
of size $r$,
which is explicitly known, and
$\sigma_{dP}$ is a model dipole-proton cross-section.

There are two
important features of $\sigma_{dP}$ that are incorporated in the GBW model.
Firstly, instead of taking $\sigma_{dP}$ to depend independently on $r$ and $W$
($\gamma^{\ast}p$ CMS energy) it is assumed that $\sigma_{dP}$
depends on Bjorken $x$
\begin{equation}
x=Q^{2}/{(Q^{2}+W^{2}-M_{p}^{2})} \label{Bjx}%
\end{equation}
where $M_{p}$ stands for the proton mass. Furthermore, in the GBW model $\sigma_{dP}$
depends on $x$ through the saturation scale $Q_{\text{s}}$ (\ref{defQsat}), namely:
\begin{equation}
\sigma_{dP}(r^{2},x)=\sigma_{dP}(r^{2}Q_{\text{s}}^{2}(x)).
\end{equation}
Given the fact that for massless quarks distribution amplitude $\psi$ is a
function of a product of $r$ and $Q$:%
\begin{equation}
|\psi(r,Q^{2})|^{2}=Q^{2}|\tilde{\psi}(rQ)|^{2}%
\end{equation}
it follows that%
\begin{equation}
\sigma_{\gamma^{\ast}p}=\text{function}(Q^{2}/Q_{\text{s}}^{2}(x)) \label{GS1}%
\end{equation}
a phenomenon known as geometrical scaling \cite{Stasto:2000er}.

Let us finish this section by recalling that the dipole-proton cross-section
$\sigma_{dP}$ is in the GBW model related to the unintegrated gluon
distribution $\varphi(x.k_{\text{T}}^{2})$ \cite{NikZakh,GolecBiernat:1998js}:%
\begin{equation}
\sigma_{dp}(x,r)=\frac{4\pi^{2}}{3}%
{\displaystyle\int}
\frac{dk_{\text{T}}^{2}}{k_{\text{T}}^{2}}\left(  1-J_{0}(rk_{\text{T}%
})\right)  \alpha_{\text{s}}\varphi(x,k_{\text{T}}^{2})
\end{equation}
where%
\begin{equation}
xG(x,Q^{2})=%
{\displaystyle\int\limits^{Q^{2}}}
dk_{\text{T}}^{2}\,\varphi(x,k_{\text{T}}^{2}). \label{glue}%
\end{equation}
The fact that $\sigma_{dp}$ cross-section depends only on the combination
$r^{2}Q_{\text{s}}^{2}(x)$ implies that%
\begin{equation}
\varphi(x,k_{\text{T}}^{2})=\varphi(k_{\text{T}}^{2}/Q_{\text{s}}^{2}(x))
\label{GSphi}%
\end{equation}
and strong coupling constant $\alpha_{\text{s}}$ should be frozen.

\section{Geometrical Scaling in pp Collisions}

\label{sectpp}

For pp collisons for low and moderate
$p_{\text{T}}$ one uses Gribov--Levin--Ryskin formula \cite{167082}%
\begin{equation}
\frac{dN}{d\eta d^{2}p_{\text{T}}}=\frac{C}{p_{\text{T}}^{2}}%
{\displaystyle\int}
d^{2}\vec{k}_{\text{T}}\,\alpha_{\text{s}}\varphi_{1}(x_{1},\vec{k}_{\text{T}%
}^{2})\varphi_{2}(x_{2},(\vec{k}-\vec{p})_{\text{T}}^{2}). \label{Nchdef}%
\end{equation}
Here $x_{1,2}$ are gluon momenta fractions
\begin{equation}
x_{1,2}=e^{\pm y}\,{p_{\text{T}}}/{W}\;\;\;{\rm with}\;\;\;W=\sqrt{s} \label{x12}%
\end{equation}
needed to produce a gluon of transverse momentum $p_{\text{T}}$ and rapidity
$y$ (or pseudorapidity $\eta$). Unfortunately formula (\ref{Nchdef}) has been
proven \cite{hep-ph/0111362} only for the scattering of a dilute system on
a dense one ({\em i.e.}
$x_1<<x_2$ or $x_2<<x_1$). Despite that, one is often forced to use (\ref{Nchdef})
in a region of  $x_1 \sim x_2$ where numerical studies show that it still works
reasonably well (see e.g. Ref.\cite{arXiv:1005.0631}).

If unintegrated gluon densities scale
according to Eq.(\ref{GSphi}) than also $dN/d\eta d^{2}p_{\text{T}}$ should
scale (provided we freeze $\alpha_{\text{s}}$ and neglect possible energy
dependence of constant $C$). Therefore geometrical scaling for the
multiplicity distribution in pp collisions
\cite{McLerran:2010ex,Praszalowicz:2011tc} states that particle spectra depend
on one scaling variable%
\begin{equation}
\tau=p_{\text{T}}^{2}/{Q_{\text{s}}^{2}(p_{\text{T}},W)} \label{taudef}%
\end{equation}
where $Q_{\text{s}}^{2}(p_{\text{T}},W)$ is the saturation scale (\ref{defQsat})
at $x_1 \sim x_2$ (\ref{x12}):
\begin{equation}
Q_{\text{s}}^{2}(p_{\text{T}},W)=Q_{0}^{2}\left(  p_{\text{T}}/%
{(W\times10^{-3})}\right)  ^{-\lambda} \label{Qsdef}%
\end{equation}
where we have neglected rapidity dependence of $x_{1,2}$.
Factor $10^{-3}$ corresponds to the choice of the energy scale
(arbitrary at this moment $x_0$ in Eq.(\ref{defQsat})).
Hence
\begin{equation}
N_{\text{ch}}(W,p_{\text{T}})\overset{\text{def}}{=}\left.  \frac
{dN_{\text{ch}}}{d\eta d^{2}p_{\text{T}}}\right\vert _{W}=\frac{1}{Q_{0}^{2}%
}F(\tau) \label{GSpp}%
\end{equation}
with $Q_{0}\sim1$ GeV. Here $F(\tau)$ is a universal function of  $\tau$.

In order to examine the quality of geometrical scaling in pp collisions
in Ref.\cite{Praszalowicz:2011rm} we have considered ratios
$R_{W_{1}/W_{2}}$
\begin{equation}
R_{W_{1}/W_{2}}(p_{\text{T}})\overset{\text{def}}{=}
\frac{N_{\text{ch}}(W_{1},p_{\text{T}}%
)}{N_{\text{ch}}(W_{2},p_{\text{T}})}.%
\end{equation}

Here we shall discuss another way of establishing geometrical scaling,
at least qualitatively. Note that if at two different energies $W_{1}$ and
$W_{2}$ multiplicity distributions are equal %
\begin{equation}
N_{\text{ch}}(W_1,{p_{\,\text{T}}^{(1)}})=N_{\text{ch}}(W_2,{p_{\,\text{T}}^{(2)}})
\label{Nchequality}%
\end{equation}
then this means that they correspond to the same value of variable $\tau
$ (\ref{taudef}). As a consequence%
\begin{equation}
p_{\,\text{T}}^{(1)\,2}\left(  {p_{\,\text{T}}^{(1)}}/{W_{1}}\right)
^{\lambda}=p_{\,\text{T}}^{(2)\,2}\left( {p_{\,\text{T}}^{(2)}}/{W_{2}%
}\right)  ^{\lambda} \label{pTW}%
\end{equation}
for constant $\lambda$. Equation (\ref{pTW}) implies%
\begin{equation}
S^{p_{\text{T}}}_{W_{1}/W_{2}}\overset{\text{def}}{=}{p_{\,\text{T}}^{(1)}}/{p_{\,\text{T}%
}^{(2)}}=\left(  {W_{1}}/{W_{2}}\right)  ^{\frac{\lambda}{2+\lambda}}.
\label{ratiospT}%
\end{equation}

\begin{figure}[h]
\centering
\includegraphics[scale=0.60]{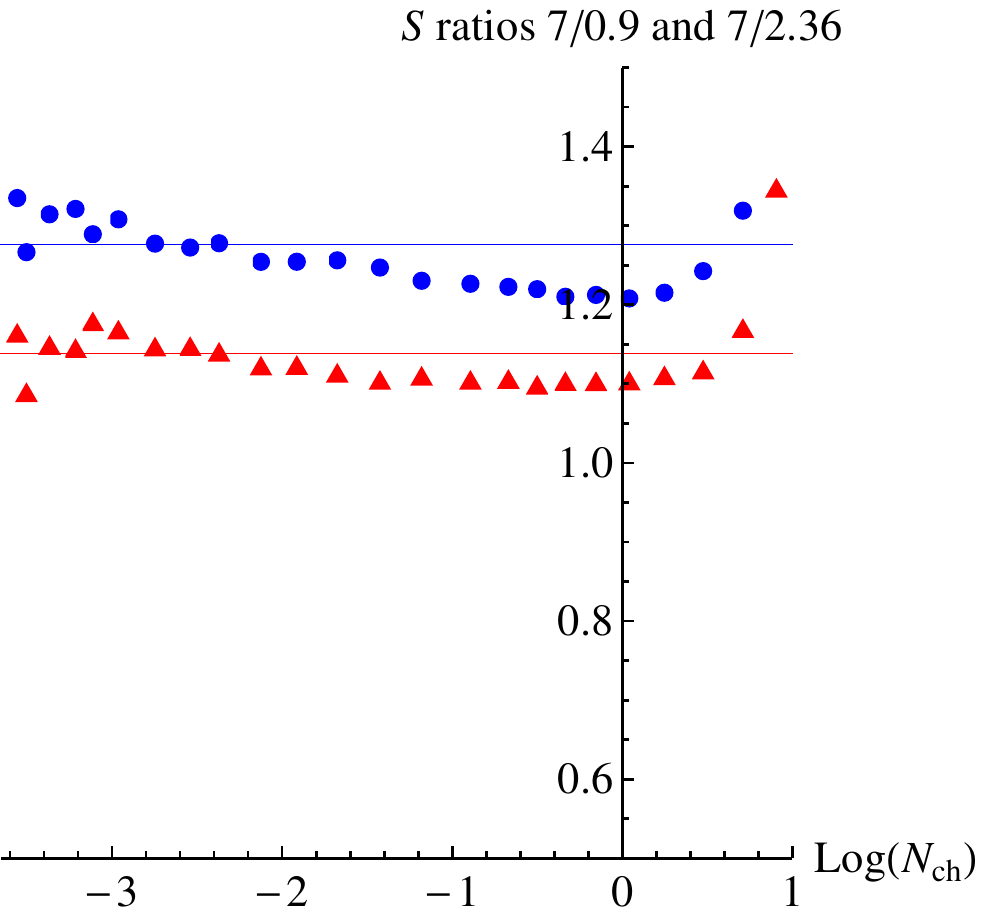}\;\;
\includegraphics[scale=0.648]{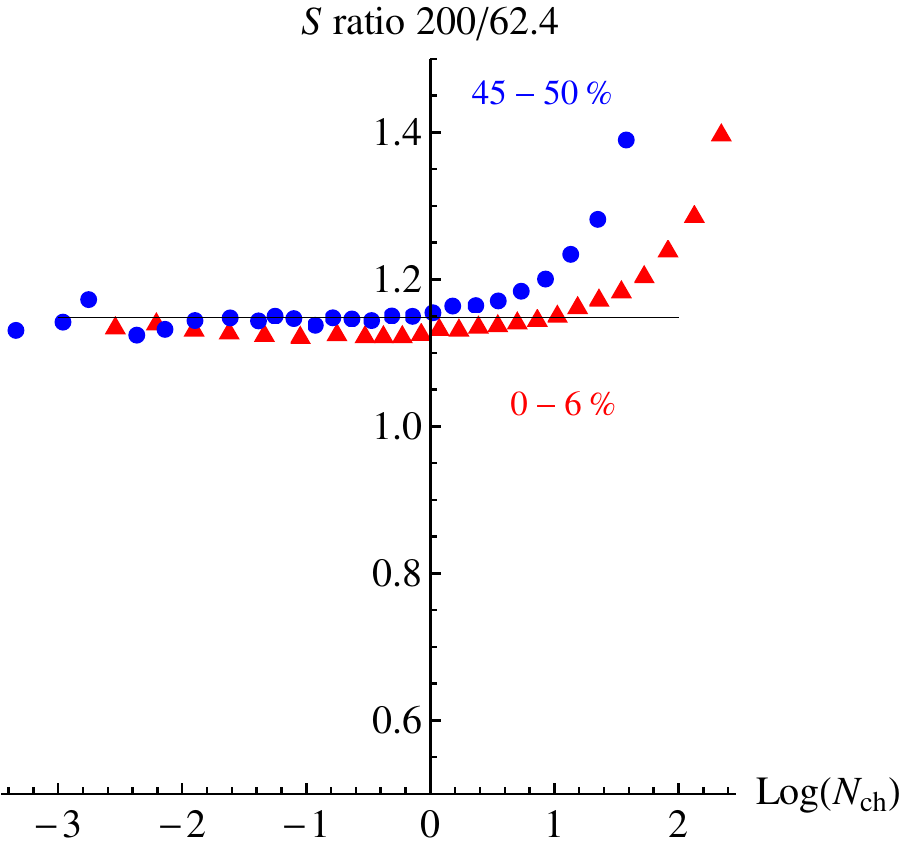}
\caption{Plots of $S^{p_{\text{T}}}_{W_{1}/W_{2}}$. Left panel: CMS pp spectra for $W_{1}=7$~TeV and $W_{2}=0.9$~TeV (blue full circles)
and 2.36~TeV (red triangles) in function of $\log(N_{\mathrm{ch}})$.
Right panel: PHOBOS Au-Au spectra
for $W_{1}=200$~GeV and $W_{2}=62.4$~GeV in
for two centrality classes: $0-5\%$
(red triangles) and $45-50\%$ (blue full circles) in function of
$\log(N_{\mathrm{ch}})$.
Straight lines correspond to the r.h.s. of Eq.~(\ref{ratiospT}) for $\lambda=0.27$.}%
\label{fig:pTratios}%
\end{figure}

Ratios $S^{p_{\text{T}}}_{W_{1}/W_{2}}$ for pp non-single diffractive spectra
measured by the CMS \cite{Khachatryan:2010xs} collaboration at the LHC
are plotted in the left panel of Fig.~\ref{fig:pTratios}
together with the straight horizontal lines corresponding to the r.h.s. of
Eq.(\ref{ratiospT}) for $\lambda=0.27$. We see approximate constancy of
$S_{W_{1}/W_{2}}^{p_{\text{T}}}$ over the wide range of $N_{\text{ch}}$. A
small rise of $S_{W_{1}/W_{2}}^{p_{\text{T}}}$ with decreasing $N_{\text{ch}}$
corresponds to the residual $p_{\text{T}}$-dependence of the exponent $\lambda$
\cite{Praszalowicz:2011tc}.

We would like to advertise this method as a simple way of looking for GS in
the $p_{\text{T}}$ spectra in various reactions. An obvious advantage is that
it is very easy to do: it requires only to interpolate one spectrum (here we
have done this for $W_{1}=7$ TeV). An obvious disadvantage consists in the
fact that it is very difficult to attribute sensible error to the ratios
$S^{p_{\text{T}}}_{W_{1}/W_{2}}$, so for quantitative purposes it is better to
consider ratios $R_{W_{1}/W_{2}}$.

\section{Onset of Geometrical Scaling in Heavy Ion Collisions}

\label{sectHI}

We have already briefly discussed the onset of GS in heavy ion collisions in
Ref.~\cite{Praszalowicz:2011rm}. This work is still in progress.
If GS in HI collisions holds then, analogously to equation (\ref{ratiospT}),
we can form ratios of $p_{\text{T}}$'s corresponding to the same multiplicity
(in the same centrality class $c$) at two different energies $W_{1,2}$.
In what follows we will neglect possible dependence of the saturation
scale upon participant density  \cite{Kharzeev:2000ph} (for
review see Ref.~\cite{McLerran:2010ub}).

Let us concentrate on the PHOBOS data for Au-Au \cite{Back:2004ra,Alver:2005nb}.
PHOBOS data cover asymmetric
rapidity region $0.2<\eta<1.4$ at two RHIC energies 64.2 and 200 GeV. The
highest centrality bin for PHPBOS is only 45--50 \%. In the right panel of
Fig.~\ref{fig:pTratios} we plot ratio $S_{200/62.4}^{p_{\text{T}}}$ for two
centrality classes: 0--6 \% and 45--50 \%, together with a line corresponding
to $\lambda=0.27$. From Fig.~\ref{fig:pTratios} we might conclude that
geometrical scaling works well also in heavy ion collisions.
More detailed  studies show, however, that dependence of the saturation scale
on the participant densities may be crucial for proper description of the data.

\section{Conclusions}

In this note we have proposed a simple procedure to look for geometrical
scaling in the $p_{\mathrm{T}}$ spectra, namely to construct ratios of
transverse momenta corresponding to the same multiplicity. Based on this
method one can qualitatively see GS in the CMS spectra \cite{Khachatryan:2010xs}
at the LHC energies and in HI collisions at RHIC \cite{Back:2004ra,Alver:2005nb}.

Many aspects of GS require further studies. Firstly, new data at higher energies
(to come) have to be examined. Secondly, more detailed analysis including identified
particles and rapidity dependence has to be performed. On theoretical side
the universal shape $F(\tau)$ has to be found and its connection to the
unintegrated gluon distribution has to be studied. That will finally lead to
perhaps the most difficult part, namely to the breaking of GS
in pp.

Even more work is needed to understand GS quantitatively for HI
collisions.
Here the main question arises why hydrodynamic evolution
preserves GS over a time of a few fermi until the final state particles are
produced. Moreover, the behavior with energy of transverse participant
densities for different centralities and for different rapidity regions
require further studies. And finally $A$ dependence of the saturation
scale has to be confirmed. Here the new LHC data for different energies and
different nuclei will open a new chapter, since the LHC energies are order
of magnitude higher than those of RHIC.

\section*{Acknowledgements} The author wants to thank the organizers
of the Low-$x$ Meeting 2011 for putting together this very successful workshop.
Special thanks
are due to Larry McLerran and Andrzej Bialas for stimulating discussions.
This work was supported in part by the Polish NCN grant 2011/01/B/ST2/00492.

\end{document}